\documentclass{aastex631}

\newcommand{\mathbold}[1]{\mbox{\boldmath $\bf#1$}}

\newcommand\mubold{{\mathbold \mu}}
\newcommand\Vbold{{\mathbold V_{\oplus,\perp}}}

\usepackage{verbatim}
\usepackage{amsmath}

\shorttitle{OB14-1760Lb: Jupiter-Sun analogue in the Galactic Bulge}
\shortauthors{Rektsini et al.}

\graphicspath{{./}}
\begin{document}

\title{OGLE-2014-BLG-1760: A Jupiter-Sun analogue residing in the Galactic Bulge}


\author[0000-0002-1530-4870]{Natalia E. Rektsini}
\affiliation{School of Natural Sciences,
University of Tasmania,
Private Bag 37 Hobart, Tasmania, 7001, Australia}
\affiliation{Sorbonne Universit\'e, CNRS, Institut d'Astrophysique de Paris, IAP, F-75014, Paris, France}

\author[0000-0003-2388-4534]{Clément Ranc}
\affiliation{Sorbonne Universit\'e, CNRS, Institut d'Astrophysique de Paris, IAP, F-75014, Paris, France}

\author{Naoki Koshimoto}
\affiliation{Department of Earth and Space Science, Graduate School of Science, Osaka University, Toyonaka, Osaka 560-0043, Japan}

\author[0000-0003-0014-3354]{Jean-Philippe Beaulieu}
\affiliation{School of Natural Sciences,
University of Tasmania,
Private Bag 37 Hobart, Tasmania, 7001, Australia}
\affiliation{Sorbonne Universit\'e, CNRS, Institut d'Astrophysique de Paris, IAP, F-75014, Paris, France}

\author{David P. Bennett}
\affiliation{Code 667, NASA Goddard Space Flight Center, Greenbelt, MD 20771, USA}
\affiliation{Department of Astronomy, University of Maryland, College Park, MD 20742, USA}

\author[0000-0003-0303-3855]{Andrew A. Cole}
\affiliation{School of Natural Sciences,
University of Tasmania,
Private Bag 37 Hobart, Tasmania, 7001, Australia}

\author{Sean K. Terry}
\affiliation{Code 667, NASA Goddard Space Flight Center, Greenbelt, MD 20771, USA}
\affiliation{Department of Astronomy, University of Maryland, College Park, MD 20742, USA}

\author{Aparna Bhattacharya}
\affiliation{Code 667, NASA Goddard Space Flight Center, Greenbelt, MD 20771, USA}
\affiliation{Department of Astronomy, University of Maryland, College Park, MD 20742, USA}

\author{Étienne Bachelet}
\affiliation{IPAC, Mail Code 100-22, Caltech, 1200 E. California Blvd., Pasadena, CA 91125, USA}

\author{Ian A. Bond}
\affiliation{Institute of Natural and Mathematical Sciences, Massey University, Auckland 0745, New Zealand}

\author{Andrzej Udalski}
\affiliation{Astronomical Observatory, University of Warsaw, Al. Ujazdowskie 4,00-478 Warszawa,Poland}

\author[0000-0001-5860-1157]{Joshua W. Blackman}
\affiliation{Physikalisches Institut, University of Bern, Gesellschaftsstrasse 6, 3012 Bern, Switzerland}

\author{Aikaterini Vandorou}
\affiliation{Code 667, NASA Goddard Space Flight Center, Greenbelt, MD 20771, USA}
\affiliation{Department of Astronomy, University of Maryland, College Park, MD 20742, USA}
\author[0009-0003-5810-1314]{Thomas J. Plunkett}
\affiliation{School of Natural Sciences,
University of Tasmania,
Private Bag 37 Hobart, Tasmania, 7001, Australia}

\author{Jean-Baptiste Marquette}
\affiliation{Laboratoire d'astrophysique de Bordeaux, Univ. Bordeaux, CNRS, B18N, alle Geoffroy Saint-Hilaire, 33615 Pessac, France}

\affiliation{Sorbonne Universit\'e, CNRS, Institut d'Astrophysique de Paris, IAP, F-75014, Paris, France}

\correspondingauthor{Natalia E. Rektsini}
\email{efstathia.rektsini@utas.edu.au}
\begin{abstract}

We present the analysis of OGLE-2014-BLG-1760, a planetary system in the galactic bulge. We combine Keck Adaptive Optics follow-up observations in $K$-band with re-reduced light curve data to confirm the source and lens star identifications and stellar types. The re-reduced MOA dataset had an important impact on the light curve model. We find the Einstein ring crossing time of the event to be $\sim$ 2.5 days shorter than previous fits, which increases the planetary mass-ratio and decreases the source angular size by a factor of 0.25.  Our OSIRIS images obtained 6 years after the peak of the event show a source-lens separation of 54.20 $\pm$ 0.23 mas, which leads to a relative proper motion of $\mu_{\rm rel}$ = 9.14 $\pm$ 0.05 mas/yr, larger than the previous light curve-only models. Our analysis shows that the event consists of a Jupiter-mass planet of $M_{\rm p}$ = 0.931 $\pm$ 0.117 $M_{\rm Jup}$ orbiting a K-dwarf star of $M_*$ =  0.803 $\pm$ 0.097 $M_{\odot}$ with a $K$-magnitude of $K_{\rm L}$
= 18.30 $\pm$ 0.05, located in the galactic bulge or bar. We also attempt to constrain the source properties using the source angular size $\theta_*$ and $K$-magnitude. Our results favor the scenario of the source being a younger star in the galactic disk, behind the galactic center, but future multicolor observations are needed to constrain the source and thus the lens properties. 

\end{abstract}


\section{Introduction} 
\label{sec:intro}

Gravitational microlensing is a highly specialized technique to discover exoplanetary systems. More than 200 exoplanets have been detected via this technique since the idea that observations towards the Galactic center can lead to exoplanet detections \citep{mao1991}. Contrary to most planet detection methods, microlensing is independent of the host-star mass and brightness. This allows us to detect planetary companions orbiting any type of massive object, whether stellar, substellar, or a compact remnant, in the Milky Way \citep{gaudi2012}. A unique opportunity given by microlensing is to be able to probe the populations of planets in the far disk and the bulge of the Galaxy, in the line of sight to the Galactic center. This can provide new insights about planetary system formation theory as a function of metallicity and environment.

Observing and then modeling the light curve of a microlensing event will provide accurate mass ratios between the star and its planet, and its projected separation in units of the angular Einstein ring radius. To learn about the physics of planetary systems, it is important to know with accuracy the physical parameters, mass and distance of the host star and its planetary companion. Often, these physical parameters are derived from Bayesian analysis with a Galactic model, which can lead to typical uncertainties of 50\% or more. In addition, this method can be insufficient for resolving degenerate or ambiguous models \citep{gould2006, choi2012, yee2021}. 
This means that we need to seek additional sources of information to constrain the physical parameters of the system.  
 
Once such constraint comes from the change of alignment between the observer, the lens and the source because of the Earth's orbit, leading to deformation of the light curve due to parallax effects. Measuring the microlensing parallax can lead to a strong constraint for the mass and distance of the lens. Unfortunately, this requires events to have quite long timescales of a few months, which is not always the case. A second approach is to use the source angular size. If the source transits the caustic, we are able to calculate an estimate of the Einstein ring radius, which can be translated into a mass-distance relation for the lens.  Note that from the light curve alone, we only measure the source size and in most cases this measurement provides only a weak constraint for the Einstein ring radius. 

A third route is to re-observe the microlensing system in the decade that follows the source-lens alignment using high angular resolution techniques. Previous works \citep{Batista2015ApJ,bennett2015} have shown that using 10-m class telescopes equipped with adaptive optics, and/or the {\it Hubble Space Telescope} can help us resolve the source and lens. This provides a constraint on the flux of the lens and allows to measure the amplitude and direction of the source-lens relative proper motion with high precision. Measuring the relative proper motion also places a strong constraint on the angular Einstein ring radius and the microlensing parallax. This means that it is possible to combine three different mass and distance constraints on the lens which can often lead to mass measurements of up to $\sim 10 \% $ precision \citep{Bhattacharya_2018,rektsini24}.

High angular resolution observations have the further use that their analysis forms part of the preparation strategy for the {\it{Nancy Grace Roman Space Telescope}}. {\it{Roman}} will be one of the first space missions to carry a microlensing survey (Roman Galactic Bulge Time Domain Survey, \cite{gaudi2024}). The survey is expected to discover more than 1400 bound planets \citep{penny19} and on the order of a thousand unbound planets \citep{Johnson2020A,sumi23} via the microlensing technique. The ultimate goal is to maximize the number of planet detections while ensuring the precise mass and distance measurements for a large fraction of them during the five-year survey. Several studies investigating the systematics and best approach to the analyses have already been completed \citep{koshimoto17, Vandorou_2020,  Bennett_2020,Blackman2021, Terry_2021,terry2022, bh2023}, covering a large variety of microlensing events.  

Here we re-visit the microlensing event OGLE-2014-BLG-1760, one of the most distant planetary systems discovered to date. In the discovery paper, \cite{Bh16} show that this system is most likely to be a gas giant planet orbiting a G-, K-, or M-dwarf star near the Galactic bulge; this is conditional on the source star being located either close to the Galactic center or further away in the Galactic disk. We use Keck adaptive optics observations of the source and lens 5.94 years following the event in order to constrain the source-lens relative proper motion and flux ratio and finally deduce the precise mass and distance of the lens as well as the source color. We confirm that the planetary system is a cold Jupiter analog in the Galactic bulge. 

The paper is organized as follows. Section~\ref{sec:previous} describes the microlensing event OGLE-2014-BLG-1760 and the conclusions drawn in the original detection paper. Section~\ref{sec:follow-up} presents our Keck high angular resolution follow-up observations and their analysis. In Sections~3.2-3.5 we describe our point spread function (PSF) fitting procedure, fitting the source and lens and measuring their separation and flux ratio. In Section~\ref{lc-model} we present a new analysis of the light curve model with and without the Keck constraints and in Section~\ref{lens-char} we show our results for the planetary system. In Section~\ref{source} we study the source properties and finally in Section~\ref{conclusion} we summarize our analysis and discuss our conclusions.

\section{OGLE-2014-BLG-1760: the event} \label{sec:previous}

The event was first announced as OGLE-2014-BLG-1760 by the Optical Gravitational Lensing Experiment (OGLE) Early Warning System \citep{udalski1994,udalski2004optical} on 2014 August 22. The Microlensing Observations in Astrophysics (MOA) collaboration \citep{Bond2001} announced the event as MOA-2014-BLG-547 on 2014 August 31. The light curve, including a planetary cusp feature \citep[Fig. 2 from ][]{Bh16}, was well-covered by the microlensing follow-up groups RoboNet \citep{tsapras09} and $\mu$FUN \citep{gould2010}. Observations from RoboNet were conducted with 1 m robotic telescopes at Sutherland, South Africa and at Siding Spring, Australia in the Sloan $i$ band. Observations from the $\mu$FUN group were made with the 1.3m SMARTS CTIO telescope in $I$-, $V$- and $H$-bands but they weren't used for the light curve analysis. In this work we decided not to use the $\mu$FUN as their contribution doesn't affect the outcome of this study. The equatorial coordinates of the event are R.A.$ = 17^h57^m38^s.16 $, dec.\ $= -28^{\circ}57^{'}43{''}.37$ (J2000.0) and the Galactic coordinates  are $l = 1\fdg 3186$, $b=-2\fdg2746$.

The event was first analyzed by \cite{Bh16}. The light curve is very well sampled in $I$- and $R$-bands but has only three $V$-band observations by OGLE, two at the beginning and one at the end of the significantly magnified period.
This leads to a color measurement of $(V-I)_{\rm OGLE,measured}=1.45$, with a fairly large uncertainty of $\pm$0.11. \cite{Bh16} also used the MOA $R$-band observations to calculate the (V-I) source color using the well constrained $\rm R_{MOA} - I$ color \citep{gould2010b}. This gives them $\rm (V-I)_{\rm MOA-OGLE,fitted} = 1.52 \pm 0.11$. Finally, they combine the two methods and obtain a source color of $\rm (V-I)_{\rm adopted} = 1.48 \pm 0.08$.

In \citet{Bh16}, the extinction towards this direction, as in \cite{bennett2014}, is calculated by using the centroid of the red clump in the color-magnitude diagram as a standard crayon. In the OGLE-III catalogue, this is found to be 
$\rm (V - I, I)_{RC} = (2.20, 15.84)$. Following \cite{bensby2011chemical} and \cite{nataf2013reddening} they adopted the dereddened red clump centroid as  
$\rm (V - I, I)_{RC} = (1.06, 14.39)$, which gives them the extinction to the source star residing inside the Galactic bulge to be 
$\rm (E(V-I),A_I)_{\rm RC} = (1.14, 1.45)$ which gives the intrinsic source color and magnitude of $\rm (V - I, I)_{S,O} = (0.34 \pm 0.08, 17.62 \pm 0.14)$.

The event has a quite short time scale of around 16 days, with a very faint and highly blended source star. The magnification is mostly defined by the MOA-R data since they were the only group to record the caustic cusp passage. This makes the $V$-band and the ($V-I$) source color highly unreliable. The rather blue color of the source-star that was estimated in \cite{Bh16} lead the authors to two different scenarios: the source can be a main-sequence star in the Galactic bulge, or a more luminous and bluer star in the far Galactic disk, behind the bulge. 
There are very few stars with such blue ($V-I$) color in the bulge but the authors argue that since the stellar density of the disk is much lower than that of the bulge, the source star is more likely to be located in the bulge, rather than beyond. This impacts on the interpretation of the distance and mass of the lens, and suggests that high angular resolution follow-up is needed to further understand the nature of the planetary system.

\section{Keck Adaptive Optics follow-up} \label{sec:follow-up}

We observed OGLE-2014-BLG-1760 on 2020 August 9 in $K_{\rm p}$-band, using the OSIRIS imager on the Keck I telescope, 6 years after the microlensing event. The images for this observational epoch can be found at \cite{https://doi.org/10.26135/koa3}. \cite{Bh16} estimated a geocentric source-lens relative proper motion of $\mu_{\rm rel,geo}$ = 6.55 $\pm$ 1.12 mas~yr$^{-1}$ so we expected a source-lens separation comparable to the OSIRIS image resolution. The pixel scale of the OSIRIS camera is 9.96 mas/pixel. The images that we obtained had an average point-spread function (PSF) FWHM of 54 mas and showed a clear resolution of the source and lens.  

\begin{figure*}
\plotone{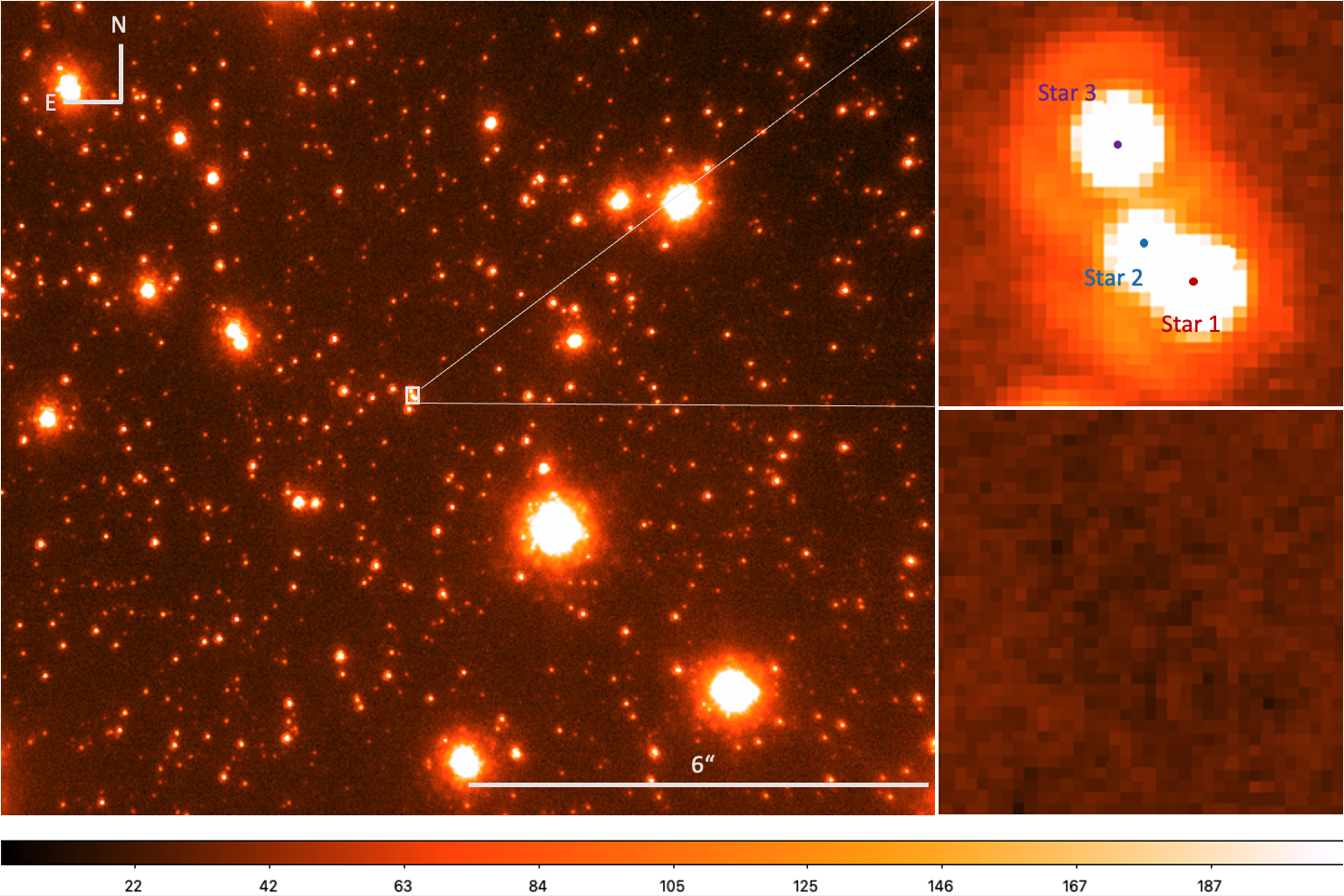}
\caption{Left panel: jack-knife stack of 20 science frames of the August 2020 Keck OSIRIS $K_{ \rm p}$ band follow up observation. Upper right panel: close-up  (0.38"$\times$ 0.38") frame of the source (star 1), the lens (star 2) and the third star (star 3), which is unrelated to the event. Lower right panel: close-up (0.38"$\times$ 0.38") of the three-star PSF fit residual using DAOPHOT.
\label{fig:s-l}}
\end{figure*}

\subsection{Analysis of the Keck images}

 We obtained 20 science images, 10 dark, 20 flat field and 10 sky frames. We used the  \texttt{KAI}\footnote{https://doi.org/10.5281/zenodo.6677744} (Keck AO Imaging) data reduction pipeline \citep{jessica_lu_2022_6522913}   to correct for geometric distortion and instrumental signatures of the OSIRIS camera, differential atmospheric refraction and cosmic ray masking and produce a co-added science frame as shown in Figure~\ref{fig:s-l}.

\subsubsection{Precise position of the source star} \label{source_id}
Inspection of the Keck image revealed four stars close to the expected position of the source. (IB) identified the event's position by comparing the MOA star field with our OSIRIS stacked frame and found the expected position of the target to be close to star-1 as shown in Figure~\ref{fig:s-l}. In order to double-check the target identification we also use the OGLE-IV catalog. (AU) linearly transformed coordinates of stars from the OGLE survey to the OSIRIS image, with residuals being at most at the 20-30 mas (2-3 OSIRIS pixels).
Then, he recalculated the centroid of OGLE-2014-BLG-1760 based on 10
subtracted images taken at maximum magnification and transferred
its position to the OSIRIS grid. The transformed centroid is located in the center of star-1. 
The accuracy of the centroid is $\approx$1~OSIRIS pixel (0.03 OGLE-IV pixel), so it can be neglected compared to the transformation accuracy. The latter is about 2-3 OSIRIS pixels. In combination with the expected source magnitude in $K$-band from the light curve analysis, we conclude that the centroid of the magnified source is conclusively identified with the star-1 and not the fainter North East companion some 5.4 pixels away (labeled star-2 in Fig.~\ref{fig:s-l}).

 \subsubsection{Photometric Calibration of the Keck images}

We use standard techniques \citep[e.g.,][]{beaulieu2016revisiting,beaulieu2018combining} to calibrate the OSIRIS stacked image. We calibrated against 2MASS $K$-band magnitudes from the VVV survey \citep{minniti2010vista}. Thirteen isolated stars are common to VVV and OSIRIS. We estimated the uncertainty in our calibration to be 3\%. We performed aperture photometry on the Keck OSIRIS frame, and measured the combined flux of star-1 and star-2 to be $\rm K_{\rm S+L} = 16.92 \pm 0.06$. 

\subsubsection{Preparing the Keck images for Jack-knife approach}
We also used the Jack-knife routine as in previous studies \citep{Bhattacharya_2021,Terry_2021} to produce a set of 20 images of 19 science co-added frames each, in order to obtain a  distribution of the quality of the science frames we are using. This is important for quantifying the Strehl ratio and PSF full-width-half-maximum uncertainties caused by atmospheric turbulence for each scientific frame. The PSF of each stacked frame of N$-$1 science images produced an average FWHM of 54 $\pm$ 0.2 mas ensuring the good quality of all the science frames used. Finally, to measure the source and lens precise positions and magnitudes, we perform PSF-fitting photometry on each of the 20 stacked science frames. We use DAOPHOT \citep{Stetson_1987} to model the sky background, obtain initial magnitude estimates by aperture photometry, and finally construct an appropriate PSF model. 

\subsection{PSF model}
We began by running the standard DAOPHOT FIND and PHOT routines to fit the sky background and estimate instrumental magnitudes for each star. We restricted the search for good PSF stars to those with instrumental magnitude no more than 0.6~mag fainter than star-1. In addition, we chose to search only within a radius of 400 pixels around the target to ensure that all chosen stars will have the same PSF and also to eliminate those that are too close to the image edges. We found 47 stars that match the magnitude and distance criteria. We identified the best candidates for the construction of our PSF model by visual inspection. The PSF stars are required to be well-isolated stars, to prevent light blending from neighbors, which will degrade the PSF model. They also need to match the ideal shape of a point source at the location of the target.

Finally, the candidates must not be saturated.

Out of the 47 candidates we found 11 that matched all the criteria and we used them to construct empirical PSF models using the DAOPHOT PSF routine.

As a last quality check, we tested different empirical PSF models with different combinations of the 11 star candidates. We rejected two stars due to their bad residual quality  and we have tested 4 different PSF models using different combinations of the 9 remaining stars. Finally, we identified the 6 stars that produce the empirical PSF model with the lowest Chi value, which represents the root-mean-square of the residuals that are left after the fitting expressed as a fraction of the peak height of the analytic function. This means that the lowest the Chi value, the larger the percentage of the stellar profiles that can be described by the empirical model.

The four models gave Chi values of 0.0595, 0.0545 and 0.0491; we chose the stars that gave the latest value.

\begin{figure}
\plotone{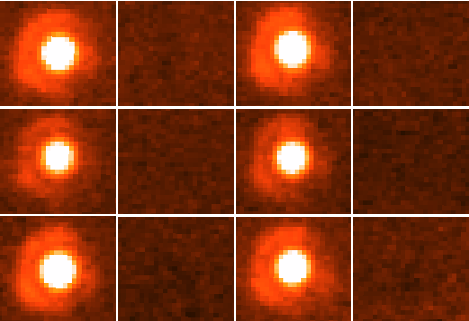}
\caption{Close-up  (0".2 $\times$ 0".2) frames of the 6 stars used to construct the PSF model (right) and their residuals, after 
profile-fitting photometry subtraction (left).
\label{fig:psf}}
\end{figure}

All 6 stars used to construct the empirical PSF model and their residuals after stellar profile fitting subtraction are shown in Figure~\ref{fig:psf}.

\subsection{Source-lens flux ratio and separation}\label{flux}

As explained in Section \ref{source_id}, we identified the source star with star-1 in Fig.~\ref{fig:s-l}, and the lens star with star-2. 

We used DAOPHOT's ALLSTAR routine to fit the positions and instrumental magnitudes for all the stars of the frame. A third star in close proximity to the target, which is also identified by the routine, contributes additional flux at the position of the target. We therefore run a 3-star PSF model to ensure simultaneous fit of all three stars. Our 3-star model produced a clean residual as shown in Fig.~\ref{fig:s-l} and the properties of each star are shown in Table \ref{tab:daophtot}. The third star is more than 100 mas away from the target, meaning it did not contribute to the lensing event. 

Finally, we used DAOPHOT$\_$MCMC \citep{Terry_2021}, a modified version of the DAOPHOT algorithm that contains an MCMC (Markov Chain Monte Carlo) routine, to produce a posterior distribution. We used the validated 3-star PSF model with DAOPHOT$\_$MCMC on each of the 20 jack-knife frames and derived the best-fit values for the source and lens separation and flux ratio. We calculated the MCMC mean and root mean square errors for all 20 frames, and also estimated the uncertainties using the jack-knife method 
\citep{Tukey1958}: 
\begin{equation}
    \sigma_x = \sqrt{ \frac{N-1}{N}\sum(x_i - \bar{x})^2     }
\end{equation}
where $N$ is the number of frames, $x_i$ is the parameter value for the $i$-th jack-knife stacked image and $\bar{x}$ is the mean value of the parameter from the jack-knife images. Our final uncertainties are the jack-knife and MCMC errors added
in quadrature as presented in Table \ref{tab:table1}.

Our estimate for the source and lens separation is 54.203 $\pm$ 0.290 mas with a flux ratio of 0.388 $\pm$ 0.011. Combining the flux ratio with the blending flux of the target in $K$-band we found the separate magnitudes of each star to be, 
$K_{ \rm star1}$ = 17.28 $\pm$ 0.050 and
$K_{ \rm star2}$ = 18.30 $\pm$ 0.054 respectively. 
In Table \ref{tab:flux} we present the summary of the source+lens, source and lens magnitudes in $K$-band as retrieved from our OSIRIS frames.

\begin{deluxetable*}{lccccccc}[htb!]
\tablewidth{20cm}
\tablecaption{DAOPHOT results for the 3-star PSF fit for the 2020 OSIRIS images \label{tab:daophtot}}
\tablehead{
 \colhead{Component}& \colhead{Coordinates}  & \colhead{F$_i$/F$_{total}$} & \colhead{s$_{12}$ (mas)} & \colhead{s$_{13}$ (mas)}  & \colhead{$\chi^2$/ dof} }
 \startdata
    $Star_1$ &    [1163.52,1157.06] &    0.4247 \\
    $Star_2$ &    [1169.41,1158.82] &    0.1661 \\
    $Star_3$ &    [1156.55,1169.41] &    0.4091 \\
    $Model$ &    {       } &    {       } &54.25 & 141.23 &       872.64$/$855 \\
    \hline
 \enddata
\tablecomments{The pixel coordinates, total flux ratio for each star, the star1-star2  and star1-star3 separation and $\chi^2$ values for the three-star PSF fitting model. The values are shown for only one image combination of N$-$1 stacked frames.}
\end{deluxetable*}

\begin{deluxetable*}{ccccc}[htb!]
\tablewidth{20cm}
\tablecaption{MCMC and Jack-knife results for the 2020 Osiris images}
\label{tab:table1}
\tablehead{
\colhead{Parameter} & \colhead{Median} & \colhead{MCMC rms} & \colhead{Jackknife rms} & \colhead{MCMC + JK rms}}
\startdata
    Separation (mas) & 54.200  & $\pm$ 0.232  &    $\pm$      0.174
    &    $\pm$ 0.290\\
      $\mu_{\rm rel,HE}$(mas~yr$^{-1}$) & 7.869 & $\pm$ 0.022 &  $\pm$ 0.052
      & $\pm$ 0.057\\
      $\mu_{\rm rel,HN}$(mas~yr$^{-1}$) & 4.641 & $\pm$ 0.014 &  $\pm$ 0.055
      & $\pm$ 0.057\\
      $\mu_{\rm rel,H}$(mas~yr$^{-1}$) & 9.144 & $\pm$ 0.058 &  $\pm$ 0.024
      & $\pm$ 0.063\\
      $\mu_{\rm rel,G}$(mas~yr$^{-1}$) & 9.131 & $\pm$ 0.094 &  $\pm$ 0.059
      & $\pm$ 0.111\\
      flux ratio   & 0.388 & $\pm$ 0.009 & $\pm$ 0.005
      & $\pm$ 0.011\\
      \hline
\enddata
\end{deluxetable*}

\begin{deluxetable}{ccc}[htb!]
\tablewidth{20cm}
\tablecaption{  $K$-band flux values}
\label{tab:flux}
\tablehead{
\colhead{Parameter} & \colhead{value}}
\startdata
      $K_{ \rm S+L}$   & 16.920 $\pm$ 0.050 \\
      $K_{ \rm source}$   &17.276  $\pm$ 0.050\\
      $K_{\rm lens}$   & 18.302 $\pm$ 0.054 \\
      \hline
\enddata
\end{deluxetable}

\subsection{Source and lens relative proper motion}\label{murel}

The source and lens separation measured in the Keck images leads to a heliocentric relative proper motion of $\mu_{\rm rel,H}$ = 9.14 $\pm$0.06 mas/yr as shown in Table \ref{tab:table1}. Meanwhile, the light curve modeling code uses relative proper motions in the geocentric frame, thus we must convert it from heliocentric into geocentric coordinates to compare the measured separation to the light curve fitting parameters and finally constrain the mass and distance of the lens system. 

To do this we use the relations from \cite{Dong_2009}: 

\begin{equation}
    \mubold_{\rm rel,geo}  = \mubold_{\rm rel,helio} - \Delta\mubold
\end{equation}
with 
\begin{equation}
    \Delta\mubold = \frac{\pi_{\rm rel} V_{\oplus,\perp}}{AU} = (\frac{1}{D_L} - \frac{1}{D_S})\Vbold
    \label{dm}
\end{equation}

where  $\Vbold$ represents the Earth's orbital velocity projected on the sky at the celestial coordinates of the lens at peak magnification, $D_{\rm S}$ is the distance to the source and $D_{\rm L}$ is the distance to the lens. We found the velocity expressed in north and east coordinates to be $\rm \Vbold $= ($-$2.76, 8.2) km~s$^{-1}$ at HJD'= 6905.767. We also used the distance to the lens estimated by \cite{Bh16}. Finally, we use the distance to the source D$_{\rm S}$ = 8.69 $\pm$1.50 kpc obtained from the Galactic model \citep{koshimoto_gal_mod} that we run 
using the \cite{Bh16} light curve model results. This gives a geocentric relative proper motion of 
$\rm \mubold_{\rm rel,geo}$ = 9.14 $\pm$0.06 mas~yr$^{-1}$. This is much higher than the predicted value from \cite{Bh16}, who found $\rm \mubold_{\rm rel,geo}$ = 6.55 $\pm$1.12 mas~yr$^{-1}$. This discrepancy can be explained by the difference between the source crossing times t$_{*}$ found by our light curve model (t$_{*}$ = 0.027 days) and by \cite{Bh16}, who found t$_{*}$ = 0.04 days.

\section{Light curve Fitting} \label{lc-model}

In order to determine the precise properties of the OGLE-2014-BLG-1760 planetary system we combine the parameters estimated by the light curve best-fit model with the AO follow-up measurements. To do this, we follow the method of \citet{bennett24,rektsini24,terry24}. 
We begin by re-fitting the light curve using the imaged-centered ray shooting method \citep{bennett96} and \citep{bennett-himag} with the results from \cite{Bh16} as initial guess. Then, we incorporate the high angular resolution results to constrain our new light curve best-fit model.
The use of AO constraints help us ensure that all the accepted light curve models will be in agreement with the lens properties determined by the high angular resolution study. 

We use a modified version of the light curve modeling code \citep{bennett96, bennett-himag}, named \texttt{eesunhong}\footnote{\url{https://github.com/golmschenk/eesunhong}}, 
in honor of the original co-author of the code \citep{rhie_phystoday,rhie_obit}. 
The code combines the microlensing event datasets with the relative proper motion and lens flux obtained by the AO follow-up analysis and uses them to constrain both the light curve parameters and the lens properties. Due to the tight relation between the relative proper motion and the microlensing parallax, the code is capable of finding the microlensing parallax even in cases when it is not observed in the light curve. In order to fit the microlensing parallax we need to also include the distance to the source $D_{\rm S}$ as a fit parameter. This is because the parallax vector is parallel to the relative proper motion in the inertial geocentric frame $\mubold_{\rm rel,geo}$, while the Keck images measure $\mubold_{\rm rel,helio}$ in the heliocentric frame. The conversion between $\mubold_{\rm rel,geo}$ and $\mubold_{\rm rel,helio}$ requires $D_{\rm S}$. Finally, the code uses a Markov Chain Monte Carlo (MCMC) algorithm with a Metropolis Hastings sampler to inspect the posterior distributions of both the light curve fitting parameters and the physical parameters of the  planetary system.
A detailed description of this code and the required parameters is included in \cite{bennett24}.

Here we illustrate the parameters of the light curve fit and present the results from models both with and without the use of the AO follow-up constraints, in addition to an extended study of the angular source radius.

\subsection{Light curve model parameters}

There are three parameters that apply to both a single lens and an N-lens microlensing event. These are t$_E$, the Einstein crossing time (length of the event), t$_0$, the moment of the minimum separation between source and lens, and u$_0$, the impact parameter that defines the minimum source-lens separation in units of the angular Einstein ring radius. In cases where the source star crosses a caustic or a cusp of a multiple lens, finite source effects may be observed in the light curve, giving t$_*$, the crossing time of the source radius. We can also estimate the angular source size $\theta_*$ using the \cite{Boyajian_2014} surface brightness relations for main sequence stars: 
\begin{equation}\label{th_s}
    log(2\theta_*) = 0.5014 + 0.419(V-I)_{\rm S,0}
    - 0.2 I_{\rm S,0}
\end{equation}
where ($V-I)_{\rm S,0}$ is the dereddened color of the source and $I_{\rm S,0}$ the dereddened source magnitude in $I$ band. There are three additional parameters that define a binary lens system: the planet-host star mass ratio $q$, their projected separation $s$ calculated in angular Einstein radius units and finally the angle $\alpha$ between their separation vector and the source trajectory. Finally, we fit two additional observational parameters, the flux of the source during the magnification F$_s$ and the blend flux F$_b$ before and after the event. The blending flux can contain the unmagnified flux of the source star as well as the lens flux in addition to other star fluxes located in close proximity to the source star. The final time-varying magnification that defined the microlensing event is expressed as F(t) = A(t)F$_s$ + F$_b$.

\subsection{Light curve Best-fit Model} \label{best_fit}

\begin{figure*}
\plotone{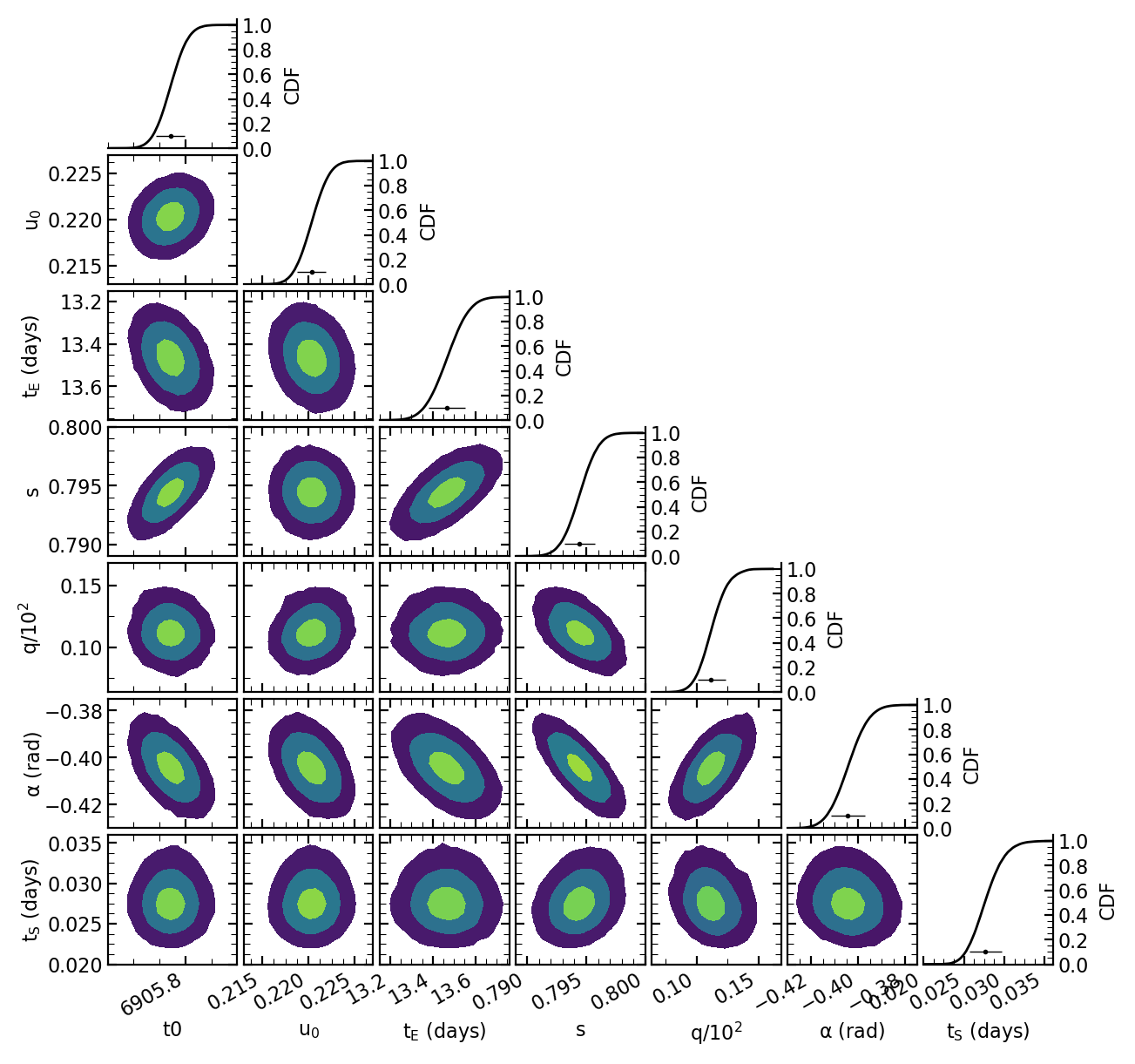}
\caption{The marginalized posterior distributions of the new light curve best-fit model. On the diagonal we show the one-dimensional cumulative density function of each parameter. The 68.3$\%$ (1$\sigma$), 95.5$\%$ (2$\sigma$) and 99.7$\%$ (3$\sigma$) confidence intervals are represented by dark violet, median blue and light green respectively.
\label{fig:corner}}
\end{figure*}

\begin{figure*}
\plotone{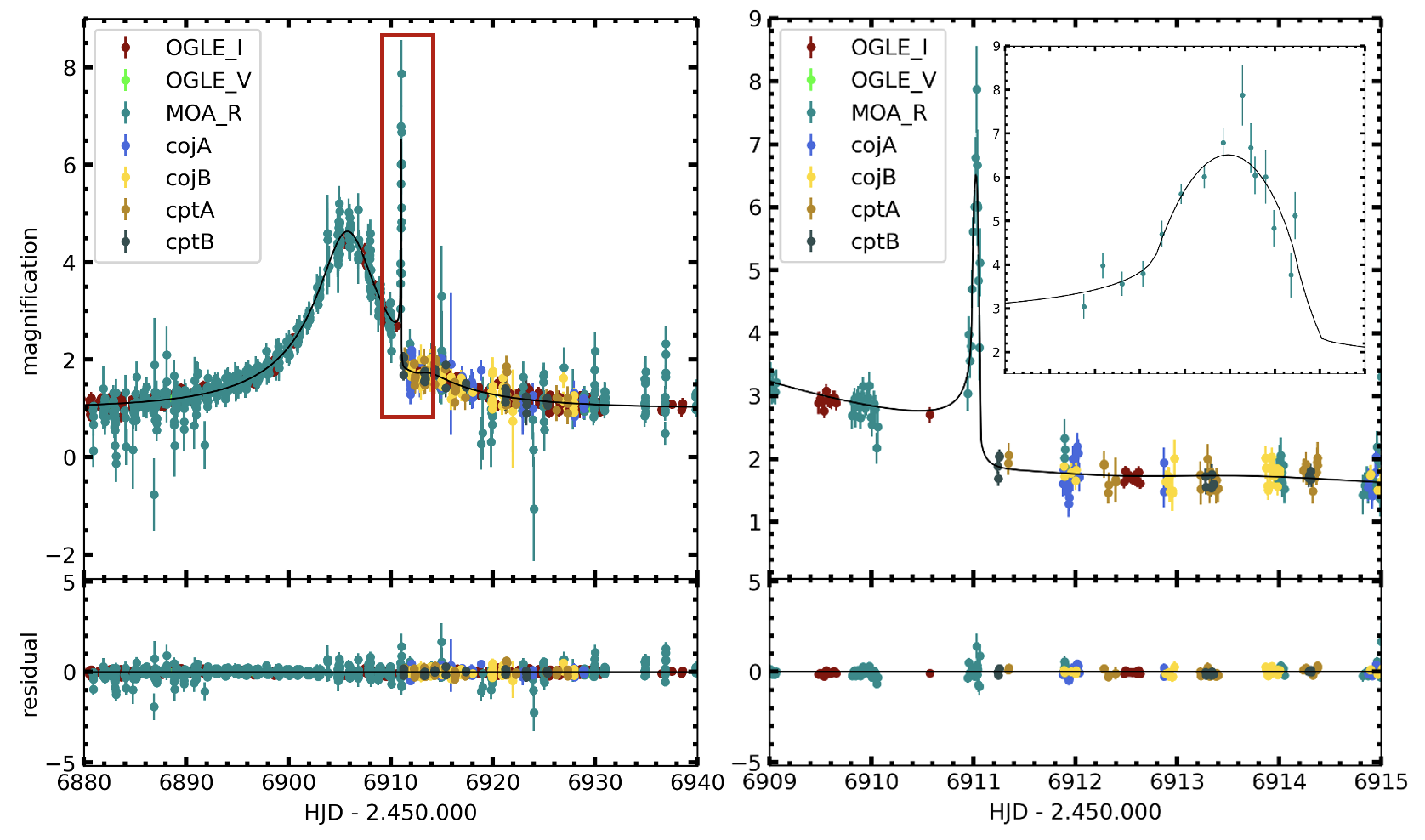}
\caption{Best-fit model of the light curve of OGLE-2014-BLG-1760. The 2L1S best-fit model is indicated by the black curve. The bottom pannel shows the residual from the best-fit model and the OGLE, MOA and RoboNet data. The right panel presents the enlargement of the caustic-crossing and the cusp of the light curve and the right-up panel shows a close-up of the anomaly. The figures were produced using the software described in \cite{moana}.
\label{fig:l-c}}
\end{figure*}

We begin by fitting the light curve data without the constraints from the high angular resolution images, using the results from \cite{Bh16} as initial guess for the two lens-one source (2L1S) fit. We use the same datasets as in \cite{Bh16} (discussed in Section \ref{sec:previous}), but we use a re-reduction of the MOA dataset as described in \cite{bond2017lowest}. Our MOA photometry has improved systematics after detrending to remove correlations with seeing and hour angle. The use of the re-reduced MOA dataset had an important impact in our new light curve model, as shown in Column 1 of Table~\ref{tab:lc_fit}.
We already observe  differences between our best-fit model and the results presented in the discovery paper. We attribute these differences to the improved MOA photometry. We present the posterior distributions of our new light curve model in Figure~\ref{fig:corner} and the 2L1S light curve fit in Figure~\ref{fig:l-c}.

\begin{figure*}
    \plotone{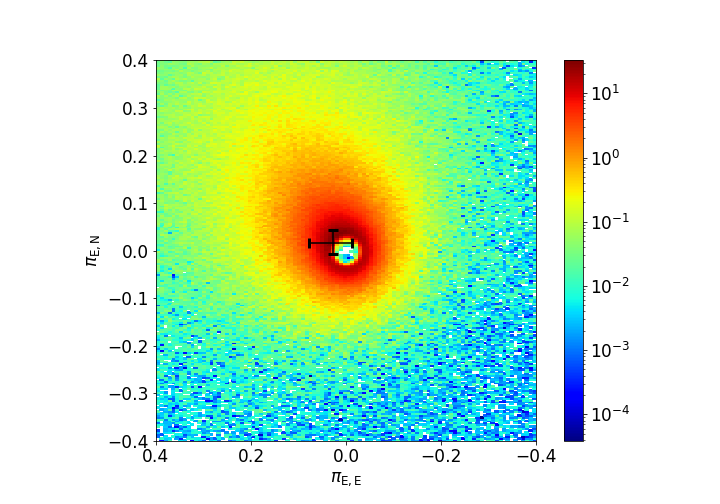}
    \caption{Two-dimensional parallax distribution based on the Galactic model (\texttt{genulens}). The color-scale shows the relative probability, the black cross indicates the microlensing parallax coordinates predicted using our (AO) Keck constraints.}
\label{fig:pie}
\end{figure*}

We proceeded to refit the light curve including the high angular resolution follow-up results, using the modified version of the \texttt{eesunhong} code. The use of the relative proper motion and lens flux measurements provide stronger constraints on the best-fit models and ensures consistent estimates of the lens and source properties. It also allows us to fit the microlensing parallax even if this is not observed during the microlensing event, when including the distance to the source $D_{\rm S}$ as a fit parameter. Since neither the distance to the source nor the microlensing parallax are directly observed for this event, we use the priors obtained from the \cite{koshimoto_gal_mod} Galactic model, using \texttt{genulens}\footnote{https://doi.org/10.5281/zenodo.6869520} \citep{naoki_koshimoto_2022_6869520}. We use $I_{\rm S}$ and ($V-I$)$_{\rm S}$ values from \citep{Bh16} in the Galactic model and obtain $D_{\rm S}$ = (8.69 $\pm$1.56) kpc for the distance to the source and ($\pi_{\rm E,E}$, $\pi_{\rm E,N}$) = [(3.579, 2.591) $\pm$ (7.321, 6.774)]$\times 10^{-2}$ respectively for the east and north parallax components. A probability distribution of the priors of the parallax components is shown in Figure~\ref{fig:pie}. The best-fit light curve model parameters with the AO constraints are presented in Column 2 of Table~\ref{tab:lc_fit}. We can see that the two sets of parameters produced for the light curve models, with and without AO constraints, are in perfect agreement.

We find the Einstein crossing time t$_E$ to be 2.40 days shorter than the one reported by \cite{Bh16}, which has a significant impact on the mass ratio $q$, which is 28\% larger than the previous value. We find a separation $s$ of 0.7943 angular Einstein radius, which confirms that s$<$1. We also find a 22\% larger impact parameter u$_0$, which is largely based on the MOA re-reduction, since the MOA-$R$ band dominates the peak of the light curve. We also report significant changes in the finite source effects, where we find the source crossing time t$_*$ to be 0.93$\times10^{-2}$ days smaller than the one reported in \cite{Bh16}. We find a microlensing parallax of $\pi_E$ = 0.052 $\pm$ 0.005 which is close to the value estimated by the Galactic model ($\pi_E$ = 0.111 $\pm$ 0.087), also shown in Figure \ref{fig:pie}, and we confirm that \cite{Bh16} correctly rejected the large, doubtful parallax value of $\pi_E$ = 5.86 that was inferred from the original light curve.

Finally, we use the same calibration relations as in \cite{Bh16} to convert OGLE-IV magnitudes to OGLE-III catalog Cousins I and Johnson V magnitudes. We observe small differences in the source I and V brightness, where we find a 0.32 less bright source in $V$ band and 0.28 difference in $I$-band. This implies a source color ($V-I$)$_{\rm S}$ = 1.40 $\pm$ 0.06. 

\subsection{Source angular size}

Regarding the source angular radius $\theta_*$, we compare the values estimated via our AO follow-up constraints and the source color and brightness. We combine the Einstein crossing time and source crossing time from the new light curve model with the constrained angular Einstein ring radius from the AO follow-up results and we can calculate $\theta_*$ using the relation : 
\begin{equation}
    \theta_* = \theta_{E}\times\frac{t_*}{t_E}
\end{equation}
where $ \theta_{E}$ = 0.336 $\pm$ 0.002 mas, as shown in Table \ref{tab:lenspar}. We find $\theta_*$ = 0.673 $\pm$0.004 microarcseconds ($\mu$as), very close to the value of 0.657 $\pm$0.011~$\mu$as predicted by \cite{Bh16}. 
In addition, we also estimate the source angular radius $\theta_*$ using Eq.\ref{th_s} for $I_{\rm S,0}$ = 17.34 $\pm$ 0.02 and ($V-I$)$_{\rm S,0}$ = 0.26 $\pm$ 0.06, finding $\theta_*$ = 0.693 $\pm$0.004 $\mu$as. 

The small difference of 0.02 $\mu$as between the two $\theta_*$ values, which is less than 2$\sigma$, confirms that the source color must be quite blue and hints towards the probability that the source star is located in the far Galactic disk, beyond the Galactic center and outside the bulge.

\begin{deluxetable*}{cccc}[htb!]
\tablewidth{20cm}
\tablecaption{Light curve best-fit model parameters. We show the MCMC mean values and 1$\sigma$ results for the best-fit obtained using only the light curve data (Column 1), the light curve data and the constraints derived by our 2020 Keck follow-up images (Column 2) and the results presented by \cite{Bh16} in the discovery paper (Column 3).}
\label{tab:lc_fit}
\tablehead{
\colhead{Parameter} & \colhead{MCMC (lc)} &
\colhead{MCMC (lc + AO)} &\colhead{Bhattacharya+16}}
\startdata
      $t_E$ (days) & 13.50 $\pm$ 0.40  &    13.47 $\pm$ 0.08  &15.87$\pm$      0.41\\
      $t_0$(HJD') & 6905.767$\pm$ 0.029 &
      6905.770$\pm$ 0.027 & 6905.856$\pm$ 0.026\\
      $u_0$ & 0.2199 $\pm$ 0.0106 &
      0.2203 $\pm$ 0.0014 & 0.1806 $\pm$ 0.0074 \\
      s   & 0.7946 $\pm$ 0.0061 &
      0.7944 $\pm$ 0.0012 & 0.8269 $\pm$ 0.0047 \\
      $\theta$ (rad) & -0.4029 $\pm$0.0106 &
      -0.4052 $\pm$0.0071 & -0.3977 $\pm$0.0086 \\
      q $\times 10^{-4}$ & 11.61 $\pm$1.27 &
      11.06 $\pm$1.09 & 8.64 $\pm$ 0.89 \\
      $t_*$ (days) & 0.0274$\pm$0.0023 &
      0.0273$\pm$0.0017 & 0.0366$\pm$0.0044 \\
      $I_s$ & 18.85$\pm$0.06 &
      18.79$\pm$0.05 & 19.07$\pm$  0.14 \\
      $V_s$ & 20.09$\pm$0.06 &
      20.19$\pm$0.06 & 20.51$\pm$  0.26 \\
      $\pi_{ \rm E,E}$ & -- & 0.0450 $\pm$ 0.0058 &  -- \\
      $\pi_{ \rm E,N}$ & -- & 0.0251 $\pm$ 0.0033 &  -- \\
      $\pi_E$ & -- & 0.052 $\pm$ 0.005 &  5.86 \\
      $D_S$(kpc) & -- & 7.943 $\pm$ 1.708 & --\\
      $\chi^2$ & 26179/26344 & 26180/26341 & -- \\
      \hline
\enddata
\end{deluxetable*}

\section{Lens System Properties} \label{lens-char}

\subsection{Lens Mass-Distance Relations} \label{m-d}
Planets discovered via the microlensing technique are usually located at 1-7.5~kpc distance. For this reason we characterize the microlensing planetary systems by measuring the mass and distance of the lens, in addition to the planet's orbital distance to the host-star. As we mentioned briefly in the introduction, it is possible to use empirical mass-distance relations for the lens to better constrain the physical parameters of the system.

The first method to estimate the lens properties comes from the source-lens relative proper motion. 
A precise measurement of the geocentric relative proper motion leads to the angular Einstein radius $\theta_{E}$ , since $\theta_{E}$ = $\mu_{\rm rel,geo} \times t_E$, from which we can derive the mass and distance relation:  
\begin{eqnarray}
    M_L = \frac{\theta_E^2}{\kappa \pi_{\rm rel}},
\end{eqnarray}

where $\pi_{\rm rel} = {\rm AU\,} ({D_L}^{-1} - {D_S}^{-1})$ for source and lens distances in kpc, and $\kappa = \frac{4G}{c^2 AU}$ = 8.144 mas$M_{\odot}^{-1}$.

The second relation depends on the microlensing parallax. This can be observed for long time scale events where the parallax will occur naturally due to the Earth's orbit as in \cite{koshimoto2017}. It can also be observed by simultaneous ground-based and space-based observations of microlensing events \citep{udalski2015,street2016}. The measurement of the microlensing parallax can provide a strong relation between the distance and the mass of the lens due to its definition: 
\begin{equation}
     \pi_E = \sqrt{\frac{\pi_{\rm rel}}{\kappa M_L}}
\end{equation}

There are very few events with precise microlensing parallax detection from the ground, so for short duration events like OGLE-2014-BLG-1760 that do not have observations from space during the lens event, the parallax mass-distance relation requires additional information such as that provided by high angular resolution flux constraints.

Finally, the measurement of the lens magnitude m$_L(\lambda)$ can be used in a mass-luminosity function for different ages and metalicities of main sequence stars \citep{delfosse2000}. 
Furthermore, we can combine the mass-luminosity relation with stellar isochrones \citep{girardi2002theoretical} of main sequence stars and construct a third mass-distance relation:  
\begin{eqnarray}
    m_L(\lambda) = 10 +5\log_{10}(D_L/1 kpc) + A_{\rm K_L}(\lambda) +\\
    \nonumber 
    \eqnum{12}
    M_{\rm isochrone}(\lambda, M_L, age, [Fe/H])
\end{eqnarray}

where $A_{\rm K_L}(\lambda)$ is the $K$-band lens extinction 
and $M_{\rm isochrone}$ is the absolute magnitude in $\lambda$ wavelength.

\begin{figure*}
\plotone{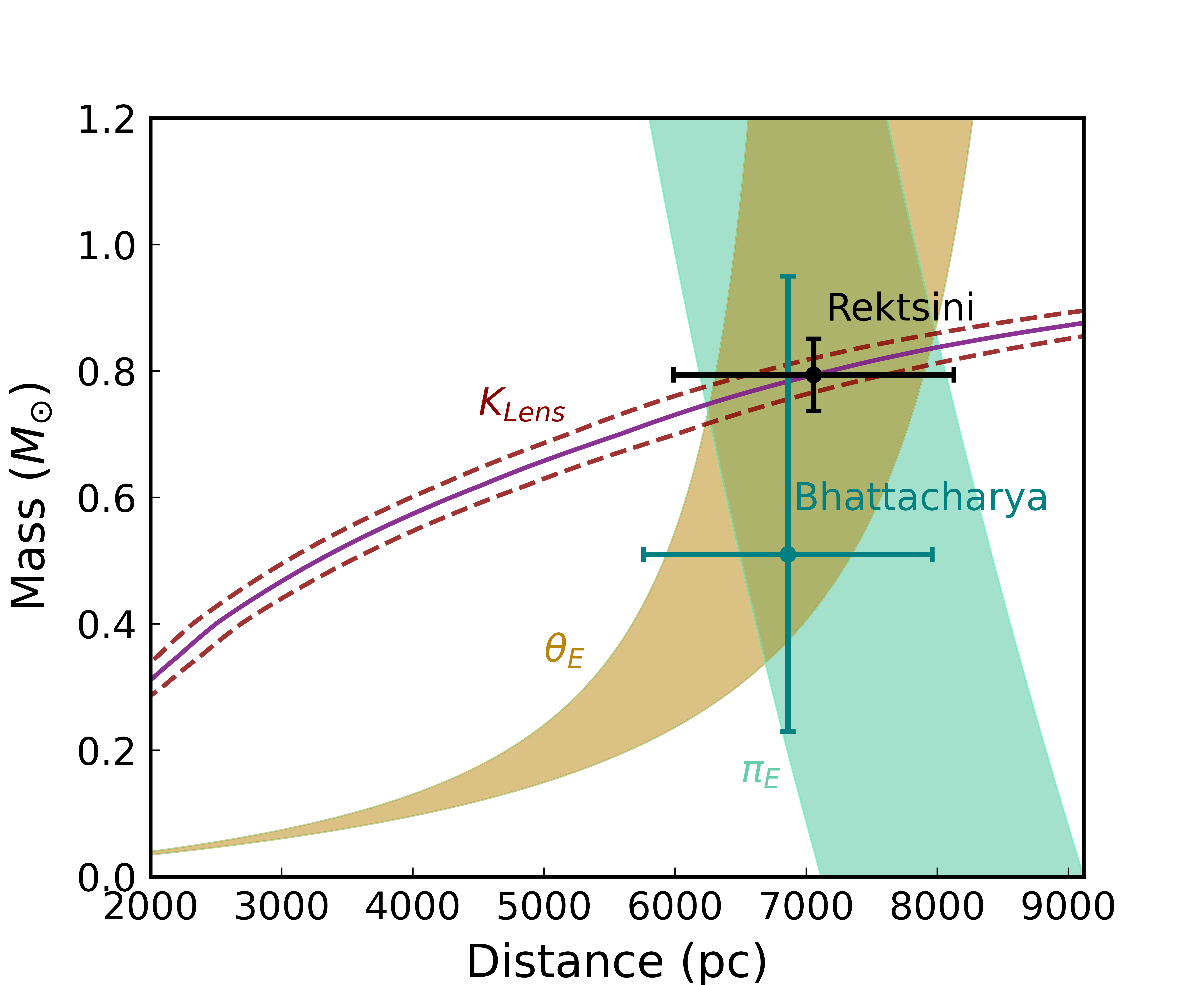}
\caption{Mass-distance estimate for the lens. The purple curve represents the constraint from the $K$-band lens flux measurement for isochrones up to 10 Gyrs, the red dotted lines represent the flux uncertainties, the gold curve shows the Einstein angular radius measurement and the aquamarine curve represents the microlensing parallax constrained by the (AO) results. The intersection between the three curves defines the estimated solution of the lens physical parameters.
}
\label{fig:M-D}
\end{figure*}

The extinction to the lens $A_{\rm K_L}(\lambda)$ is estimated by combining the extinction in $K$-band with the distance to the source and the distribution of the Galactic dust in relation to the source and lens locations. If we assume the Galactic dust distribution as an exponential in both radius and height in a disk \citep{drimmel2001three} then the extinction to the lens can be approximated as : 
\begin{equation}
\label{akl}
    A_{K_L} = \frac{1- e^{ \rm -\vert D_L(sinb) / h_{\rm dust}\vert}}{1- e^{ \rm -\vert D_{S}(sinb) / h_{\rm dust}\vert}} A_K
\end{equation}

where b = -2$\fdg2746$ is the Galactic latitude of the event, h$_{dust}$ = 0.10 $\pm$ 0.02 kpc is the dust scale height, $D_{ \rm S}$ is the distance to the source from the light curve model, and $A_{\rm K}$ is the $K$-band extinction at the distance to the source. 
In Fig. \ref{fig:M-D} we show the mass and distance graph of the lens system by combining the three mass and distance relations described here. We present the microlens parallax constrain in aquamarine, the angular Einstein radius constrain in golden. Finally we present the isochrone constrained mass-luminosity relation with purple line and red dotted lines to express the errors in the magnitude of the lens. We calculate the mass-luminosity relation for a lens extinction in $K$-band equal to $A_{\rm K_L}$ = 0.165 using the Eq. \ref{akl} and we use isochrones for ages up to 10 Gyr and metallicities within the range 0.0 $\le$ [Fe/H] $\le$ +0.3. This is consistent with the estimates of \cite{clarkson11} where they find very few stars with ages less than 5 Gyr inside the Galactic bulge but the varying metallicities observed by microlensed stars show that there should be a significant number of intermediate age (5 to 8 Gyr) stars hidden inside the bulge \citep{Bensby2013}.

\subsection{Planetary System Parameters}

By combining the light curve fitting parameters with the high-angular resolution observations using the mass-distance relations we find that the lens system is characterised by a Jupiter-Sun analogue with a planet of mass $M_p$ = 0.931 $\pm$ 0.117 $M_{\rm Jup}$ orbiting a host-star of mass $M_*$ = 0.803 $\pm$ 0.097 $M_{\odot}$ as shown in Table \ref{tab:lenspar}. We find the system residing inside the Galactic bulge at distance D$_L$ = 7.056 $\pm$ 1.468 kpc, making it one of the most distant planetary systems discovered to date. Finally, we calculate the projected separation between the planet and the host star as: 

\begin{equation}
    \alpha_{\perp} = sD_L\theta_E
\end{equation}

We find the Jupiter-mass planet to be in an orbital distance of $\alpha_{\perp}$ = 1.887 $\pm$ 0.380 au around its host-star. 

\begin{figure*}
\plotone{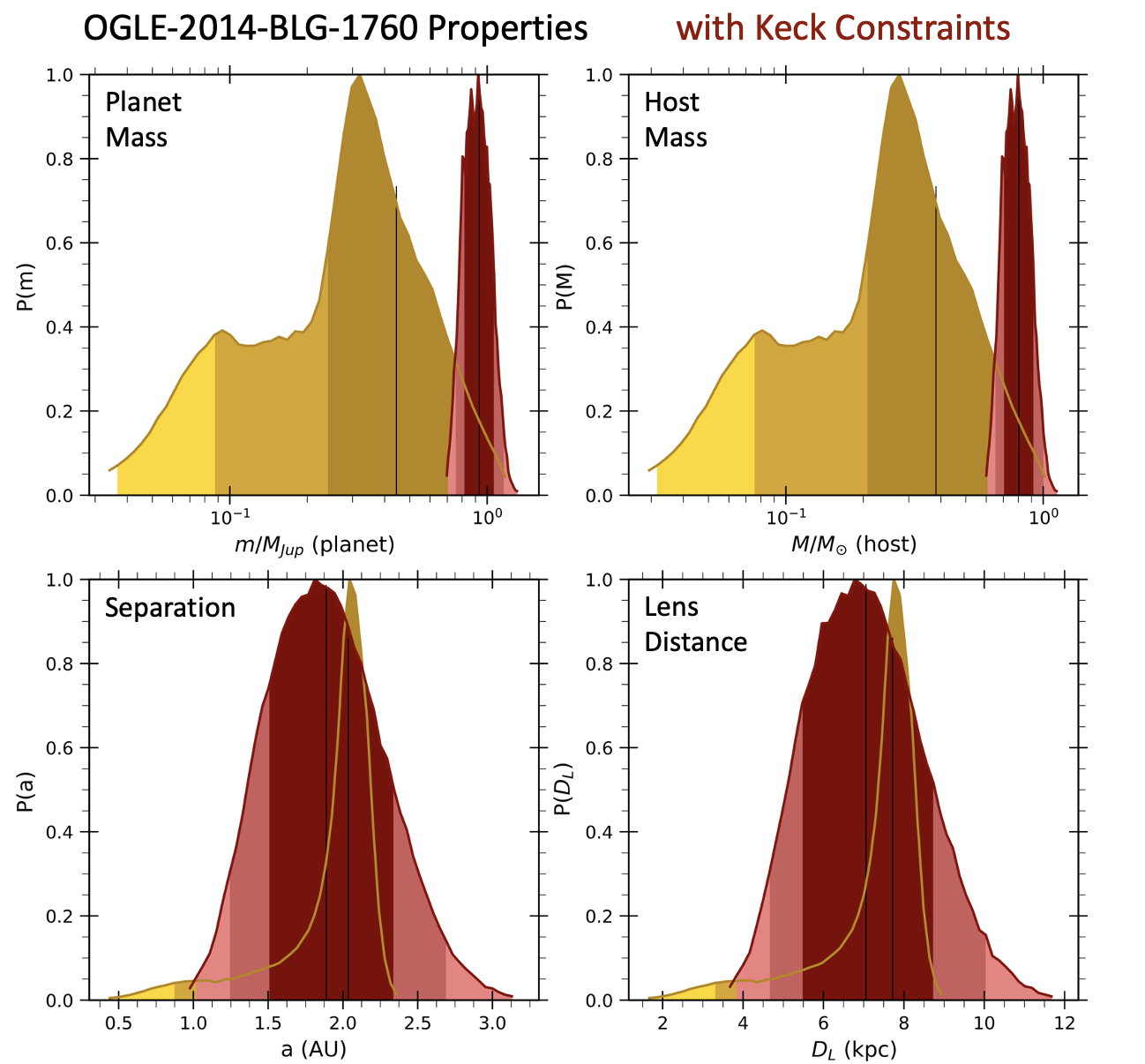}
\caption{Bayesian posterior probability distributions for the planetary companion mass, host mass, their projected separation, and the distance to the lens system are shown
with only light curve constraints in gold
and with the additional constraints from our Keck follow-up observations in red. The central 68.3$\%$ (1$\sigma$) of the distributions are
shaded in darker colors (dark red and dark gold), and the remaining central 95.4 $\%$ (3$\sigma$) of the distributions are shaded in lighter colors. The vertical black line marks the
median of the probability distribution for the respective parameters. The priors used for the Bayesian analysis are the estimates from the final light curve model.  We show that the medians of the Bayesian probability are within 2$\sigma$ of the constrained parameter distributions for the lens mass and distance.}
\label{fig:hist}
\end{figure*}

The probability distributions, in red, for the mass and distance of the lens and the planet and host-star orbital separation presented in Fig. \ref{fig:hist} are the product of the three empirical mass-distance relations combined with the constrained light curve fitting parameters. Our results show a difference between the expected values from \cite{Bh16}. This can be partly explained by the differences in the light curve model estimates and in particularly the length of the event, which affects the planet-host star mass-ratio and the source radius crossing time. 

Meanwhile, in this figure we also notice that while the Keck follow-up results manage to derive the mass of the lens very accurately, the distance to the lens seems to be better constrained by the Bayesian Galactic model, plotted in gold. 
We remind that this is a very faint and highly blended event with a poorly constrained source color which leads to significant questions about the properties of the source star. The Galactic model relies only on the light curve information and thus predicts a less massive lens, located almost inside the Galactic center and a source star located in the far-disk. This low lens mass scenario is rejected by the AO follow-up results, which revealed a much more massive lens. 
The constraints that we obtain from the Keck follow-up analysis are not affected by the size of the magnification or the quality of the light curve data, which is also the main interest of these observations, but an accurate lens distance measurement requires an accurate source distance, due to their tight relation. 
This means that we need to study further the source star properties in order to accurately constrain the planetary system parameters.

\begin{deluxetable}{ccc}[htb!]
\tablewidth{20cm}
\tablecaption{Lens Parameters Table}
\label{tab:lenspar}
\tablehead{
\colhead{Parameters} & \colhead{Units} &\colhead{Values and 1$\sigma$}}
\startdata
      Angular Einstein Radius $\theta_E$ & mas & 0.336 $\pm$ 0.002\\
      Host Mass $M_h$ & $M_{\odot}$   & 0.803 $\pm$ 0.097  \\
      Planet Mass $m_p$ & $M_{\rm Jup}$   & 0.931 $\pm$ 0.117  \\
      Lens Distance $D_L$ & kpc   & 7.056 $\pm$ 1.468 \\
      2D Separation $\alpha_{\perp}$ & au & 1.887$\pm$ 0.380\\
      \hline
\enddata
\end{deluxetable}

\section{Source star properties } \label{source}
As explained in \cite{Bh16} and showed in Fig. \ref{figure:oglev} there are no MOA-V observations for this event and the few OGLE-V points contain large uncertainties. In addition, the event is very faint and the source highly blended. This produces a lot of uncertainties in the source color and brightness. We decided to study further the source color and distance using AO follow-up constraints with the light curve model, the Bayesian analysis with the \cite{koshimoto_gal_mod} Galactic Models and the Gaussian Mixture approach.

\subsection{Source color} \label{par:source_color}
Here, we use the catalogue from \cite{surot2020mapping} to find the extinction to the source in $K$-band. For the galactic coordinates of the event $l = 1\fdg 3186$, $b=-2\fdg2746$ we find E(J-K) = 0.3350 $\pm$ 0.0100 for a distance r = 0.00072 between the \cite{surot2020mapping} grid point and the target in degrees. From \cite{nishiyama2009interstellar} we find the $A_{\rm K}$/E(J-K) = 0.494 relation and finally the $K$-band extinction as $A_{\rm K}$ = 0.165. This means that according to the \cite{Bh16} results we expect the source magnitude in $K$-band to be $K_{\rm S}$ = 17.425 $\pm$ 0.08. 

We also estimate the $K$-band of the source using the source color predicted by our new light curve model as described in Section \ref{lc-model}. For ($V-I$)$_{\rm S,0}$ = 0.26 $\pm$ 0.06 we find ($V-K$)$_{\rm S,0}$ = 0.465 using the color tables from \cite{mamajek}.This leads us to a flux value for the source of $K_S$ = 17.24 $\pm$ 0.06, which is much closer to the $K$-band magnitude measured in the Keck high angular resolution images. Finally, we use the angular source radius implied by the AO constraints in Section \ref{best_fit} and the $I_{\rm S,0}$ value from the light curve model as known parameters in Eq.\ref{th_s} and we derive a new estimate for the source color of ($V-I$)$_{\rm S,0}$ = 0.228 $\pm$ 0.070. This new color estimate aligns with the source having a very blue color. 

We derive that the $K$-band magnitude of the source for this color should be $K_S$ = 17.30 $\pm$ 0.06. A summary of the results for the angular source radius $\theta_*$, the source color ($V-I$)$_{\rm S,0}$ and source $K_{\rm S}$ magnitudes calculated for each case can be found in Table \ref{tab:source}. Our $K_{\rm S}$ source magnitude estimates from both cases are consistent with the magnitude deduced in Section \ref{flux} using the OSIRIS images and thus tend to confirm the blue color of the source star. Using the color tables from \url{https://www.pas.rochester.edu/~emamajek/EEM_dwarf_UBVIJHK_colors_Teff.txt} based on \cite{mamajek}, we estimate the source star to be an A7-dwarf with effective temperature of $T_{\rm eff}$ $\sim$ 7760 K and radius of $\sim$ 1.750 $R_{\odot}$.

\subsection{Source distance from the Galactic models }

We investigate further the galactic location of the source by running a new Galactic model using \texttt{genulens}
\citep{naoki_koshimoto_2022_6869520,koshimoto_gal_mod}. We study the source distance for all three source colors and magnitudes presented above. As shown in Table \ref{tab:source} the Bayesian analysis using the Galactic model predicts a source star located inside the Galactic bulge according to the source color adopted by \cite{Bh16} but the distance increases for the new source color, both from the new light curve model and the AO-follow up results. More precisely,  
we find that for a ($V-I$)$_{\rm S,0}$ = 0.228 $\pm$ 0.06 and $I_{\rm S,0}$ = 18.74 $\pm$ 0.06 the distance to the source is $D_{\rm S}$ = 10.321$^{\rm +1.983}_{\rm - 0.721}$ kpc, while the distance increases to 11 kpc for the source color inferred from our new light curve model, without including the AO follow-up constraints. Since the Galactic models don't  predict a source location on the far side of the bulge \citep{Bh16}, we explored the source color parameter space in order to estimate the break-point, where the model ``jumps'' to distances larger than 8~kpc. We fixed $I_{\rm S,0}$ = 18.74 $\pm$0.06, and explore the effect of increasing the color by increasing $V_{\rm S,0}$. We find that the model predicts a source closer than 9 kpc for $V_{\rm S,0}$ = 17.82 which gives a source color ($V-I$)$_{\rm S,0}$ = 0.48 $\pm$0.10. This leads to a predicted infrared magnitude $K_{\rm S}$ = 18.99 value, according to the \cite{mamajek} color tables, which is ruled out by our Keck AO imaging. In all cases, the lens distance remains around 7 kpc, demonstrating that the inferred planetary system parameters are robust to systematics in the source color.

\subsection{Source distance from the Gaussian Mixture approach}

Here we study the source and lens properties using a different approach by using the \texttt{pyLIMASS} algorithm \footnote{https://github.com/ebachelet/pyLIMA}. This tool is independent from Galactic models and uses isochrones to model the stellar properties while using a list of observables as constraints. Finally, it uses a Gaussian mixture approach to approximate the posterior distribution, sampling the most plausible physical parameters of the lens and source. Furthermore, it uses a different statistical approach to combine the source and lens fluxes from the light curve model and the AO follow-up images with the isochrones to produce the final stellar property distributions. A detailed description of the code can be found in \cite{bachelet24}. The estimates here are mostly based on the isochrones.  

We run \texttt{pyLIMASS} using as observables the $I_{\rm S}$ magnitude, the ($V-I$)$_{\rm S}$ source color, the t$_E$ Einstein time and $\rho$ = t$_*$/t$_E$ values, all parameters fitted by the light curve model, in addition to the source and lens $K$-band magnitudes measured from the AO follow-up images. Our results for the source and lens distance are compatible with the Galactic model analysis. \texttt{pyLIMASS} finds the source located further away from the Galactic bulge in all three cases, but finds a slightly smaller distance for the color from \cite{Bh16}. Furthermore, the code estimates a lens mass of 0.9 $\pm$ 0.2 $M_{\odot}$, which is in agreement with our AO$+$lc analysis and a source surface temperature of $T_{\rm eff}$ = 7900 $\pm$500 K, in agreement with the results in Section \ref{par:source_color}. 

The weight of the evidence, even though it fails to constrain the source distance value, it  strongly suggests that OGLE-2014-BLG-1760 is the first microlensing event with AO follow-up observations that confirm an early-type source star located behind the Galactic bulge.

\begin{deluxetable*}{cccccc}[htb!]
\tablewidth{20cm}
\tablecaption{  Source properties }
\label{tab:source}
\tablehead{
\colhead{Parameter}  & \colhead{$\theta_*$($\times10^{-4}$ mas) }  & \colhead{($V-I$)$_{\rm S,0}$}  & \colhead{$K_{\rm S}$}  & \colhead{$D_{\rm S}$ [kpc] (\texttt{genulens})}  & \colhead{$D_{\rm S}$ [kpc] (\texttt{pyLIMASS})}  }
\startdata
      \cite{Bh16}  & 6.57 $\pm$ 0.11    & 0.34 & 17.425 $\pm$ 0.08 & 8.69 $^{+1.449}_{- 0.761}$ & 9.59 $\pm$ 2.39\\
      light curve   & 6.932  $\pm$ 0.040 & 0.26 $\pm$ 0.06 & 17.24 $\pm$ 0.06 & 11.53 $^{+1.725}_{- 1.244}$ & 11.94 $\pm$ 1.72\\
      AO follow-up           & 6.728  $\pm$ 0.040 & 0.228 $\pm$ 0.070 & 17.30 $\pm$ 0.06 & 10.32 $^{+1.983}_{- 0.721}$ & 10.67 $\pm$ 2.32\\
      \hline
\enddata
\tablecomments{The angular size, color, magnitude in $K$-band and distance of the source predicted by the \cite{Bh16} analysis, by our updated light curve model, and by the AO follow-up constraints. The source color value from the last method gives the closer $K$-band source magnitude to $K_{\rm S}$ = 17.28 measured in the Keck images. }
\end{deluxetable*}

\begin{figure}
\plotone{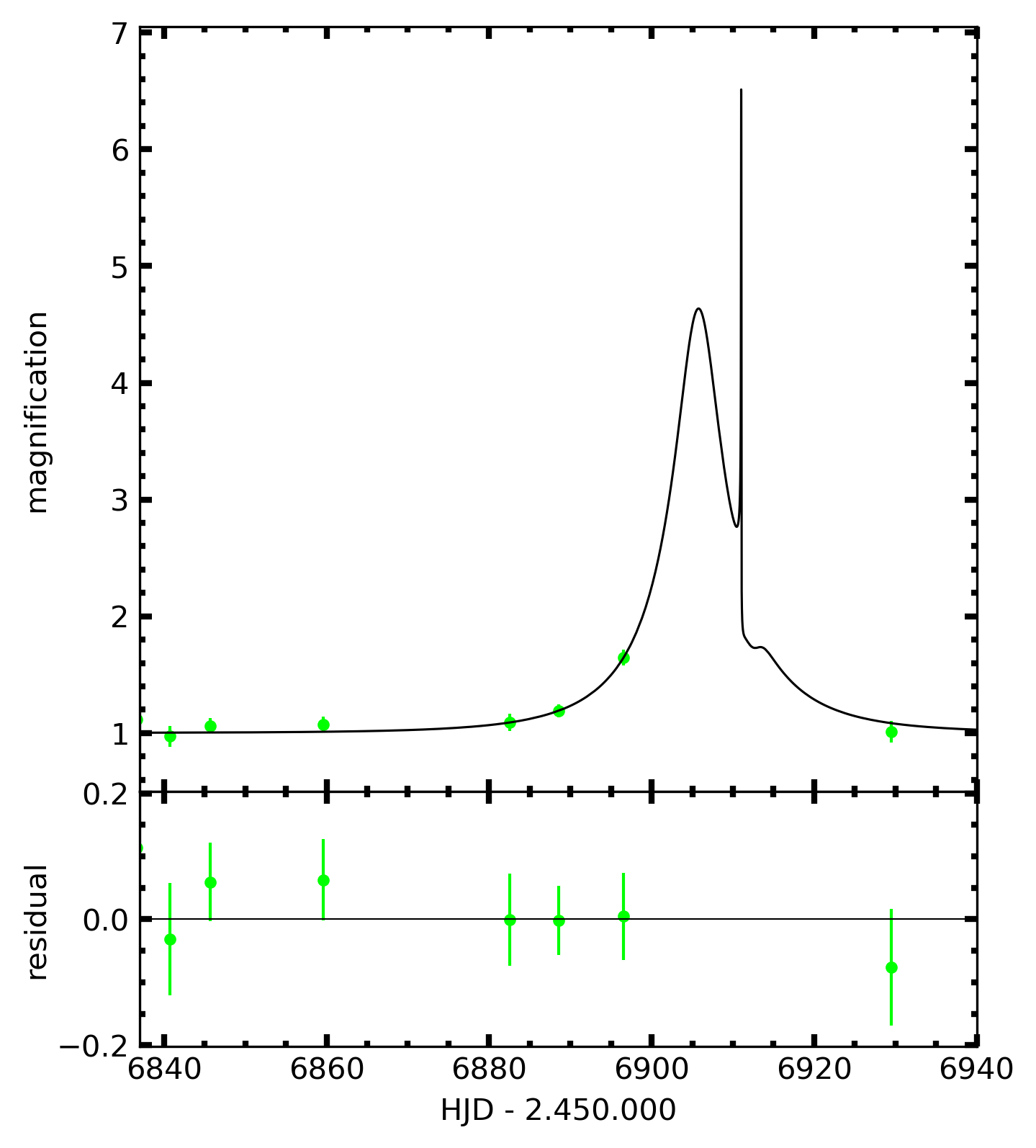} 
\caption{2L1S light curve model described by the black line with the $V$-band data. The green points showing the OGLE-V data around the microlensing event, the only $V$-band observations. At the bottom panel we show the residual from the best-fit model and the OGLE-V data.}
\label{figure:oglev}
\end{figure}

\section{Summary and Conclusion}
\label{conclusion}

Our Keck AO follow-up observations of OGLE-2014-BLG-1760 5.94 years after the microlensing peak have allowed us to separately identify the source and lens and 
reveal their characteristics. Our $K$-band observations with Keck OSIRIS imager show a lens of magnitude $K_{\rm lens}$ = 18.30 $\pm$ 0.05 and a source magnitude of $K_{\rm source}$ = 17.28 $\pm$ 0.05. The separation between source and lens was larger than predicted in the discovery paper \citep{Bh16}, leading to a 2.5 times larger relative proper motion of $\mubold_{\rm rel}$ = 9.14 $\pm$ 0.05 mas~yr$^{-1}$. This discrepancy can be resolved by the significant differences in the source size and the length of the event in our new light curve model. 

We used a modified version of the imaged-centered ray-shooting  method \citep{bennett96,bennett-himag} that uses the source and lens magnitudes and their relative proper motion as additional constrains to fit the light curve model. Furthermore, 
the use of the re-reduced R-MOA dataset has had a significant impact in our fitted light curve parameters reducing the length of the event by $\sim$ 2.4 days. This is probably due to the short Einstein time and the faintness of the event and also the fact that the  R-MOA data are dominating the light magnification. This means that even small detrending corrections can sometimes result in important alterations in the parameters that define the event. We find a larger planet host-star mass-ratio value of q = (11.06 $\pm$ 1.09)$\times  10^{-4}$, which places the event even further away from the mass-ratio break point found in the \cite{suzuki16} sample, making the event a Sun-Jupiter analogue.

The use of AO follow-up constraints for our light curve fit has permitted us to derive the microlens parallax of the event, which in return worked as an additional constrain for the lens mass and distance. Our light curve model yields a parallax value of $\pi_E$ = 0.052 $\pm$ 0.005, a value that is also confirmed by our Galactic model. Using this parallax value in addition to the lens $K$-band magnitude and relative proper motion we find the lens to be described by a Jupiter-mass planet of $M_{\rm p}$ =0.931 $\pm$ 0.117 $M_{Jup}$ orbiting an early K-dwarf star of $M_*$ =  0.803 $\pm$ 0.097 $M_{\odot}$ in $D_{\rm L}$ = 7.056 $\pm$ 1.468 kpc. This places the lens in the Galactic bulge or in the bar. The confirmation of a microlensing planet in the Galactic bulge region contradicts the claims of \cite{penny16}, there are planets in the inner Galactic bulge region. {Finally, the rather large uncertainty in our lens distance estimate is probably caused by the highly uncertain source position. This shows the importance of the precise characterization of the source properties in order to constrain the lens system physical parameters.

It is not easy to derive the source properties of this microlensing event. OGLE has obtained few measurements
in $V$-band, and all the points are at low amplification which hampered the possibility to obtain a secure $V$-band source estimate from the light curve fit (Fig. \ref{figure:oglev}). Our new light curve model also revealed a quite blue source color of ($V-I$)$_S$ = 1.4 $\pm$ 0.06 but this value is also dependent from the same $V$ data used in the discovery paper. In our attempt to validate this result we decided to estimate the source color independently from the OGLE $V$-band observations. 

We use the source magnitude measured from Keck images as reference. We derive the $K$-band of the source using the source color from the light curve model, the value is inside the 2$\sigma$ difference from the Keck $K$-band value. Then we also use the source angular radius value $\theta_*$ that we derive using the relative proper motion of the AO follow-up images and we use this value and the Eq.\ref{th_s}, with the $I_{\rm S}$ of the light curve to calculate a new estimate for the source color. Finally, we use this source color to deduce the $K$ magnitude of the source. As shown in Table \ref{tab:source}, both methods reveal $K$-band values for the source that are inside 2$\sigma$ from the initial Keck images value. 

If we estimate the source color to be ($V-I_{\rm C}$)$_{\rm S,0}$ $\sim$ 0.24 according to \cite{mamajek} we expect the source star to be an A7-dwarf of $\sim$ 1.75 $R_{\odot}$ radius. This result reinforces the case of the source being a young star residing in the Galactic disk, behind the Galactic center. Meanwhile, this scenario is also favored by our Galactic model \citep{koshimoto_gal_mod}. We used the derived $I_{\rm S}$ and source color values for all three cases presented above and find the most probable estimate of the source distance to be in  $D_{\rm S}$ = 10.321$^{+1.983}_{- 0.721}$ kpc. Finally, we also compared the source distance with the predictions derived by \texttt{pyLIMASS}. The advantage of comparing the source and lens properties with this tool is that it is independent from the Galactic models, giving more emphasis on stellar isochrone models and that it is using a different statistical method from both \texttt{eesunhong}, (MCMC algorithm with a Metropolis Hastings sampler) and \texttt{genunlens} (Bayesian analysis) to produce the stellar properties distributions. Finally, \texttt{pyLIMASS} combines information from both the light curve model and the constrained $K$-band source value from the Keck images. This means that, contrary to the Bayesian analysis with a Galactic model, it is able to reject the false smaller lens mass scenario and make more accurate predictions for the nature of the source star. Our results with \texttt{pyLIMASS} are in agreement with the lens properties inferred from the light curve model and the AO follow-up analysis, while they also favor the source star being located behind the Galactic bulge. Finally, we examine the stellar color and $I$-band properties of the stars residing inside the Galactic bulge \citep[figure 7, left panel from][]{terry20} and we find no star exhibiting similar properties in that field.

There have been studies \citep{navarro2020} as well as few previous reports of microlensing events that could contain source stars in the far disk. \cite{Shvartzvald2018} were the first to report the discovery of a giant planetary system, probably inside the Galactic bulge, and a source suffering from a severe reddening that placed it in the far disk, behind the bulge. As mentioned in the paper, the field around this event suffers from high and differential extinction which makes it very challenging to derive precise properties for the source star, but their results seem to indicate an M7 giant dwarf located in the far disk.  Similarly, \cite{Bennett2018} report the discovery of a microlensing event with an unusually red source color, which makes it impossible for the source star to be inside the Galactic bulge. They mention that one possible option would be for the source to reside in the far disk but they also favor the scenario of a lower main-sequence source star in front of the Galactic bulge. Finally, \cite{Li2019} report the discovery of a microlensing event with a variable oscillating red giant source star that could potentially be located behind the Galactic bulge.  

OGLE-2014-BLG-1760 is the only event so far with AO follow-up observations that infer a source star located in the far disk. Furthermore, the color and the $K$-band magnitude of the source indicate a young early A-dwarf star, which are very common in this Galactic region. Further observations with high angular resolution images in multiple bands will be able to confirm these results and offer a final precise measurement for the source and the lens distance.

As mentioned in \cite{Bh16} the galactic coordinates of this event are quite close to the expected observational fields of {\it{Nancy Grace Roman Space Telescope}}.
Studying and understanding further the stellar population of this field and the distance distribution of the microlensing events is very important for building efficient observational strategies for this space survey. The case of OGLE-2014-BLG-1760 represents a perfect example of the kind of planetary systems we can discover in the Galactic bulge and raises the question of the probability to observe microlensing events with source stars beyond the Galactic bulge.

\section*{Acknowledgments}
This work was supported by the University of Tasmania through the UTAS Foundation and the endowed Warren Chair in Astronomy and the ANR COLD-WORLDS (ANR-18-CE31-0002) and by NASA through grant NASA-80NSSC18K0274. This research was also supported by the Australian Government through the Australian Research Council Discovery Program (project number 200101909) grant awarded to AC and JPB. This work was also supported by CNES, focused on Euclid and Roman missions.
The Keck Telescope observations and analysis were supported by a NASA Keck PI Data Award, administered by the NASA Exoplanet Science Institute. Data presented herein were obtained at the W. M. Keck Observatory from telescope time allocated to the National Aeronautics and Space Administration through the agency’s scientific partnership with the California Institute of Technology and the University of California. The Observatory was made possible by the generous financial support of the W. M. Keck Foundation. The authors wish to recognize and acknowledge the very significant cultural role and reverence that the summit of Mauna Kea has always had within the indigenous Hawaiian community. We are most fortunate to have the opportunity to conduct observations from this mountain.

\bibliography{ob141760}{}
\bibliographystyle{aasjournal}
\end{document}